%% file: main.tex
\documentclass[a4paper]{article}
\usepackage{INTERSPEECH2022}
\usepackage{xspace}
\usepackage{todonotes}

\title{CALM: Contrastive Cross-modal Speaking Style Modeling for Expressive Text-to-Speech Synthesis}

\name{Yi Meng$^{1,\dagger}$\thanks{$^{\dagger}$Work conducted when the first author was intern at Tencent.}, Xiang Li$^1$, Zhiyong Wu$^{1,2,*}$ \thanks{$^*$Corresponding author.}, Tingtian Li$^3$, \\Zixun Sun$^3$, Xinyu Xiao$^3$, Chi Sun$^3$, Hui Zhan$^3$, Helen Meng$^{1,2}$}

\address{
  $^1$Tsinghua-CUHK Joint Research Center for Media Sciences, Technologies and Systems,\\
Shenzhen International Graduate School, Tsinghua University, Shenzhen, China\\
  $^2$Department of Systems Engineering and Engineering Management,\\
The Chinese University of Hong Kong, Hong Kong SAR, China\\
  $^3$Tencent, Shanghai, China}
\email{\{my20, xiang-li20\}@mails.tsinghua.edu.cn, \{zywu, hmmeng\}@se.cuhk.edu.hk, tingtian.li@outlook.com, \{zixunsun, lawxiao, collinsun, huizhan\}@tencent.com}

\begin{document}

\maketitle

\gdef\sys{our model\xspace}

\input{sections/abstract}
\begin{figure*}[tbp]
  \centering
  \includegraphics[width=1.01\linewidth]{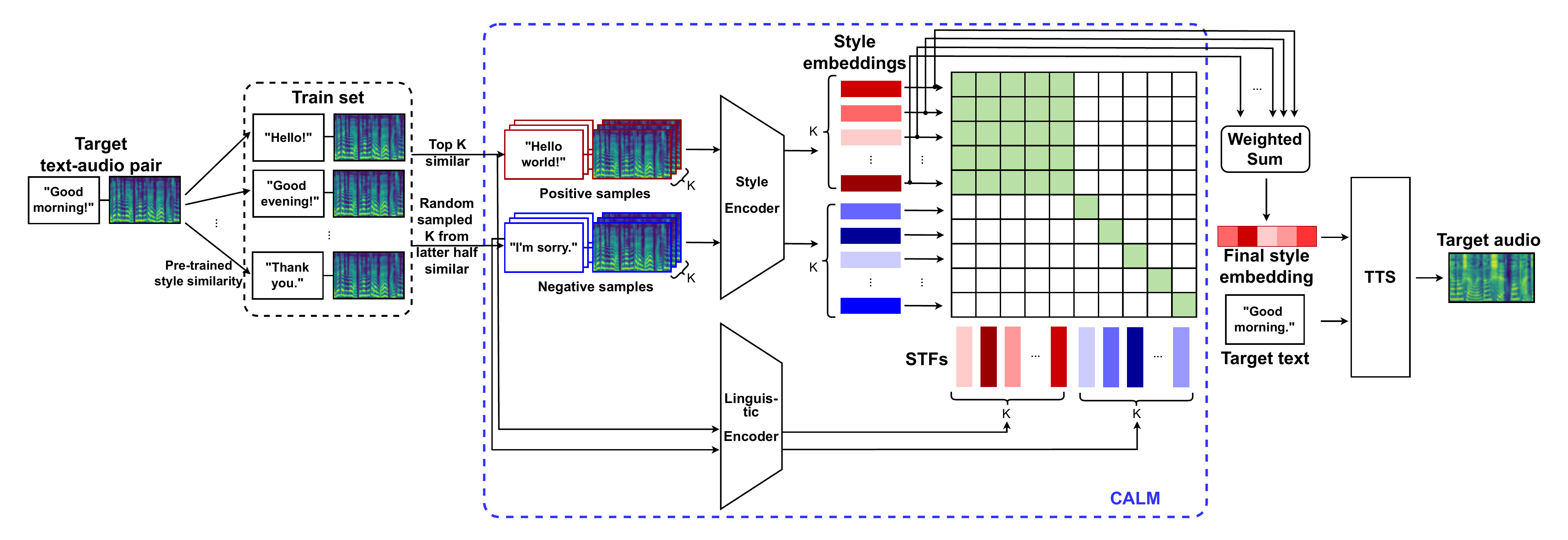}
  \caption{An overview of the training process of the proposed method}
  \label{fig:overview}
\end{figure*}
\input{sections/introduction}


\input{sections/methodology}
\input{sections/experiment}

\input{sections/conclusion}
\input{main.bbl}

\end{document}

%% file: sections/abstract.tex
\begin{abstract}
To further improve the speaking styles of synthesized speeches, current text-to-speech (TTS) synthesis systems commonly employ reference speeches to stylize their outputs instead of just the input texts. 
These reference speeches are obtained by manual selection which is resource-consuming, or selected by semantic features. 
However, semantic features contain not only style-related information, but also style irrelevant information. The information irrelevant to speaking style in the text could interfere the reference audio selection and result in improper speaking styles. 
To improve the reference selection, we propose Contrastive Acoustic-Linguistic Module (CALM) to extract the Style-related Text Feature (STF) from the text.
CALM optimizes the correlation between the speaking style embedding and the extracted STF with contrastive learning. 
Thus, a certain number of the most appropriate reference speeches for the input text are selected by retrieving the speeches with the top STF similarities. 
Then the style embeddings are weighted summarized according to their STF similarities and used to stylize the synthesized speech of TTS. 
Experiment results demonstrate the effectiveness of our proposed approach, with both objective evaluations and subjective evaluations on the speaking styles of the synthesized speeches outperform a baseline approach with semantic-feature-based reference selection.


\end{abstract}

%% file: sections/introduction.tex
\section{Introduction}
Human speeches naturally contain speaking style variations as an important component to express the intention and emotional states of the speaker,
such as intonation and speaker rates.
With the development of deep learning and speech processing \cite{shen2018natural, ze2013statistical}, many approaches have been proposed in text-to-speech (TTS) synthesis to enhance the speaking style of the synthesized speech. 
Conventionally, such enhancement is achieved by directly predicting the speaking styles from the given input texts \cite{stanton2018predicting, chien2021hierarchical}.
Variational autoencoder (VAE) \cite{kingma2013auto} is employed to model the correlation between the speaking styles and the word embeddings of given texts \cite{zhang2019learning}.
Neural network based method is further proposed and employs a CBHG \cite{wang2017tacotron} network to predict the speaking styles from the word embeddings \cite{stanton2018predicting}.

But there are differences between the linguistic information in text and the acoustic information in speaking style. Since style embedding is an entangled representation of many acoustic factors, it is difficult to predict style from text alone \cite{gong2021using}. Besides, there might be many appropriate styles for the given text, thus facing the one-to-many problem, which makes the prediction even harder \cite{gst}.
Therefore, many methods are proposed to learn the patterns of speaking styles from reference speeches.
Global style embeddings (GSE) \cite{skerry2018towards} are firstly introduced to extract the speaking style of a reference speech. 
Global style tokens (GST) \cite{gst} are further proposed to embed the GSE of the reference speech into a weighted sum of style tokens. 
These two approaches employ the ground-truth speeches as the references for training.
However, since no ground-truth reference is available during inference,
the synthesized speeches are observed to suffer from low content quality.
Multiple references TTS (MRTTS) \cite{gong2021using} is proposed and employs multiple references in both training and inference to overcome this issue and improve the robustness of TTS systems.

The reference selection in MRTTS is based on the semantic information of the text. 
However, semantic information contains not only style-related information, but also style irrelevant information. 
For example, semantic similarity is utilized in natural language processing to select the answers to a given question \cite{lan2018neural, minaee2017automatic, chergui2019integrating}, while the actual speaking styles of the question and selected answers could differ considerably. 
Take the question `How is the weather?' as an example, its semantic information is close to a typical answer like `Such a bad weather!', since their topics are both about the weather.
But the selected answer is spoken in a hateful exclamatory tone, while the question is spoken in an ordinary interrogative tone.
In this situation, MRTTS may synthesize speech with improper speaking styles.

To overcome this issue, the reference speech selection should focus on the style-related information in the text.
Inspired by Contrastive Language-Image Pre-training (CLIP) \cite{radford2021learning}, which learns perception from the supervision contained in text-image pairs, we propose a multiple reference TTS system with Contrastive Acoustic-Linguistic Module (CALM).
CALM is a contrastive-learning-based cross-modal speaking style selecting module, which learns to extract Style-related Text Features (STF) from text via audio supervision. 
CALM utilizes a style encoder to extract style embeddings from speeches, and a linguistic encoder to extract STF from the text.
During the training stage, the style encoder is first trained along with the TTS system. 
Then we jointly optimize the style encoder and linguistic encoder by pushing the STFs with similar style embedding close to each other, while pushing away STFs with dissimilar style embeddings. 
Eventually, CALM learns to map the style and text of speech into the same embedding space shared by style embeddings and STFs. 
Thus, during the inference stage, to obtain the reference speeches whose speaking styles are most appropriate for the input text, we can directly select the speeches with the most similar STFs. 
Then the speaking styles are weighted summarized to stylize the synthesized speeches of TTS. 

Experimental results demonstrate the effectiveness of our proposed approach.
Both of the objective and subjective evaluations on Mandarin emotional dataset and English audiobook dataset confirm that
CALM outperforms the semantic-feature-based baseline model on generating stylized speeches.

%% file: sections/methodology.tex
\section{Methodology}

As shown in Figure~\ref{fig:overview}, our proposed model is based on the end-to-end TTS framework FastSpeech 2 \cite{ren2020fastspeech}. Its style input comes from the Contrastive Acoustic-Linguistic Module (CALM), which consists of a linguistic encoder and a style encoder.


\subsection{Contrastive Acoustic-Linguistic Module (CALM)}
We use the linguistic encoder to extract Style-related Textual Feature (STF) from text, and the style encoder to extract style embedding from speech.
CALM maps the style and text into the same embedding space and change the distance between their corresponding embeddings by contrastive learning.
Specifically, CALM forces texts with similar speech styles (positive samples) to have closer STFs. 
Conversely, the STFs with different speech styles (negative samples) are dispersed as far as possible in the embedding space. 
During the inference stage, STF is used to guide the reference audio selection in TTS.


\subsubsection{Positive and negative samples selection} \label{selection}
To select the positive and negative samples for contrastive training, we first train the style encoder on the training set. 
Then we utilize the trained style encoder to extract style embeddings for each training audio, which is use to measure the style similarity among different samples. 
For each (speech, text) pair in the training dataset, we sort all the other pairs in the training set by their style embedding cosine similarity to the style embedding of current speech.
The top-$K$ of the sorting results are selected as positive samples, while negative samples are obtained by randomly selecting $K$ samples from the latter half of the sorted pairs.
We do not select the last $K$ samples as negative samples because the last $K$ sample is too easy for the model to distinguish.

\subsubsection{CALM training}

As shown in the blue dash line in Figure~\ref{fig:overview}, given $2K$ (speech, text) pairs as the input, CALM tries to predict a $2K \times 2K$ matrix $\boldsymbol{M}^{'}$, where the value at the position of the $i^{th}$ row and $j^{th}$ column indicates the cosine similarity between the style embedding of the $i^{th}$ speech obtained by the style encoder and the STF embedding of the $j^{th}$ text obtained by the linguistic encoder. 
The ground truth matrix $\boldsymbol{M}$ only consist of $1$ and $-1$ values: (i) $1$ for the elements whose STF embedding and style embedding are both from positive samples set or extracted from an identical pair in negative samples set; (ii) $-1$ for the others. In detail, the first $K$ (speech, text) pairs are positive samples, and the last $K$ pairs are negative samples.
The loss function $\mathcal{L}_{CALM}$ is the mean square error (MSE) between the predicted matrix $\boldsymbol{M}^{'}$ and the ground-truth matrix $\boldsymbol{M}$. 

By these settings, CALM can learn a multi-modal embedding space by jointly training a style encoder and a linguistic encoder to maximize the cosine similarity of the embeddings between the positive samples and minimize the cosine similarity between the negative samples.

\subsubsection{Model details}
The setting of the style encoder of CALM is identical to the style encoder part of a GST enhanced Fastspeech 2 \cite{gst, ren2020fastspeech}.
The linguistic encoder of CALM aims to obtain sentence level feature from text. It contains a pre-trained BERT \cite{devlin2018bert, cui2020revisiting} that takes a sentence with $L$ words as input and output word level features with shape $(L \times D)$, which will be fed into a net with a two-layers GRU \cite{cho2014learning}, dropout with rate $r$ and a linear unit to adjust the output dimension to the same dimension with style embedding. With the help of the training process mentioned above, the output text feature tend to be more related to the speech style, thus called Style-related Textual Feature (STF).

\begin{figure}[t]
  \centering
  \includegraphics[width=\columnwidth]{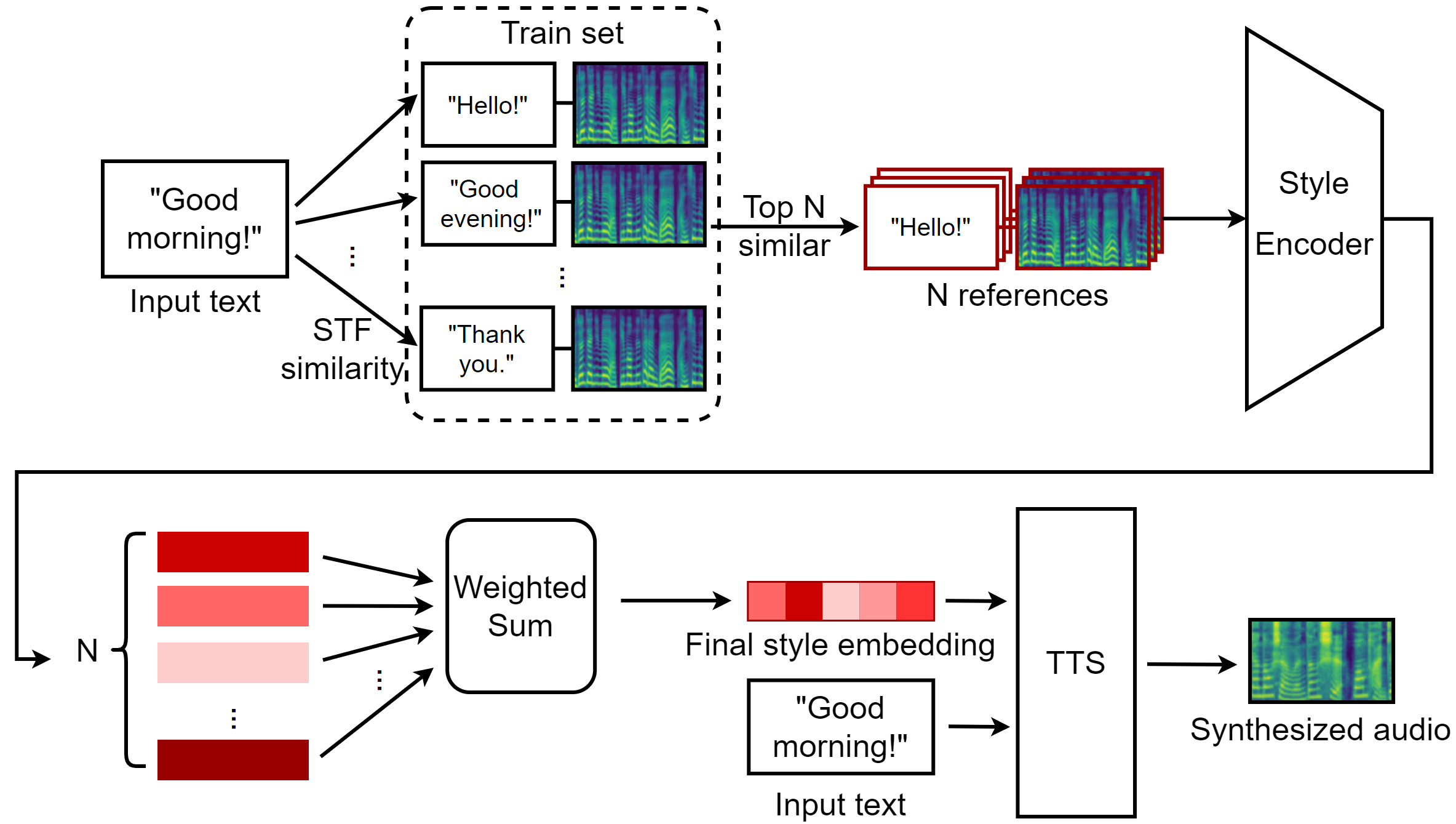}
  \caption{STF-driven multiple references selection}
  \label{fig:retrieval}
\end{figure}

\subsection{The TTS system with CALM}
\subsubsection{Training stage}
During training time, as demonstrated in Section~\ref{selection}, we firstly train a GST enhanced Fastspeech 2, namely the TTS module in Figure~\ref{fig:overview}. The pre-trained style encoder is used to extract speech styles in the training set.
After that, we get positive samples and negative samples for all utterances,
the style embeddings of the positive samples will not only be used in CALM, but also weighted summarized into the final style embedding used in TTS module. 

We define $N$ as the number of reference audios. Given $N$ style embeddings extracted from reference audios $\boldsymbol{S} = [\boldsymbol{s}_1, \boldsymbol{s}_2, ..., \boldsymbol{s}_n]$, and the STFs extract from the text of references $\boldsymbol{T} = [\boldsymbol{t}_1, \boldsymbol{t}_2, ..., \boldsymbol{t}_n]$, and the STF of the input text $\boldsymbol{t}_0$, the calculation of the final style embedding can be notated as below
\begin{equation}
  \boldsymbol{w} = softmax(\boldsymbol{T}  \boldsymbol{t}_0^T) 
\end{equation}
\begin{equation}  
  \boldsymbol{final\_style\_embedding} = \boldsymbol{w}^T  \boldsymbol{S}
\end{equation}
where $\boldsymbol{w}$ is the weights of style embeddings.
Since the positive samples in CALM training also served as reference audios to predict target audio, we set $N = K$ in training time.
Taking the final style embedding and the target text as input, the TTS module produces target audio.
We jointly train the entire TTS system from scratch with CALM.
The total loss of the whole TTS system can be defined as
\begin{equation}  
  \mathcal{L}_{total} = \mathcal{L}_{TTS} + \lambda \mathcal{L}_{CALM}
\end{equation}
where $\lambda$ is the weight of the CALM loss.
With the help of this loss function, we can jointly optimize the CALM and TTS part of the whole system.

\subsubsection{Inference stage}
As shown in Figure~\ref{fig:retrieval}, during the inference time, input text is fed into the trained linguistic encoder in CALM to get the STF. All the text of (speech, text) pairs in the training set will be fed into the linguistic encoder of CALM to get STFs. Then we calculate the STF cosine similarity with all STFs in the training set and sort them in descending order. We use top-$N$ of them as the reference audios to extract style embeddings, and weighted sumarized to get the final style embedding. Taking the final style embedding and the text as input, the TTS system can synthesize audio with corresponding speaking style. 

Unlike the training stage, the number of $N$ can vary and differ from $K$. We will discuss the influence of $N$ in Section~\ref{selN}.



%% file: sections/experiment.tex
\section{Experiments}

\subsection{Datasets}
In this paper, we use two datasets in two languages, Mandarin and English, to demonstrate the effectiveness of the proposed model. The synthesized samples are available on our demo page\footnote{https://thuhcsi.github.io/interspeech2022-CALM-tts/}.

\subsubsection{Mandarin emotional dataset} 
The first dataset is an internal single-speaker emotional Mandarin corpus, including six emotion categories (angry, fear, disgust, happy, sad, surprised). A total of 9,000 utterances is evenly divided by the six emotions. For each category of emotion, 50 utterances are randomly sampled to form the test set with a total of 300 utterances. The remaining 8,700 utterances serve as the training set. Though each utterance is labeled by emotion, we do not use the emotion label as supervision. The emotion label are only used to evaluate the performance of the reference audio selection in Section \ref{sec:obj_eval}.

\subsubsection{English audiobook dataset}
The second dataset is an open-source English audiobook dataset. The books are read by the 2013 Blizzard Challenge speaker, Catherine Byers, in an animated and emotive storytelling style. Some books contain very expressive character voices with a high dynamic range. We choose four of these audiobooks as our dataset, with 9,741 utterances in total. We randomly select 300 utterances as the testing set while keeping the rest as the training set.

\subsection{Models}
\subsubsection{Baseline}
We implement Multiple Reference TTS (MRTTS) \cite{gong2021using} as the baseline approach and replace its Tacotron-2 TTS backbone \cite{shen2018natural} with Fastspeech 2 for fairness concern. 
MRTTS selects the reference audios by semantic similarity during both the training and inference stages.
MRTTS also constrains the style embeddings by pre-trained style embeddings during the training time.
The semantic similarity is defined as the cosine similarity of two sentence-level embeddings, which is generated by a pre-trained BERT. For objective comparison, we used the same pre-trained BERT on the audiobook dataset with our proposed model, in which the output word embedding is with a dimension of 768.

\subsubsection{Proposed model details}
We utilize FastSpeech 2 \cite{ren2020fastspeech} as the backbone of our model. The style encoder is the same setting with GST \cite{gst}. In the CALM, the linguistic encoder GRU has 256 hidden units, and the drop rate is 0.2. The loss weight $\lambda$ is set to 1. The number of reference audios $K$ in training stage is 20. If not specified, the number of reference audios $N$ in inference stage is set to 20, which will be discussed in \ref{selN}. 
As for waveform generation, a well-trained HiFi-GAN \cite{kong2020hifi} is chosen as our neural vocoder.

\subsection{Evaluations}



\subsubsection{Objective evaluation}
\label{sec:obj_eval}

We first demonstrate that the STFs extracted by CALM are more related to speaking style than the semantic features extracted by BERT.
We reveal this by calculating precision on the test set of the emotional dataset.
(We can't calculate the precision for the audiobook dataset since it do not have style labels.)
A higher precision indicates that more audios with the correct emotion labels were selected. 
Specifically, given $N$ selected text-audio pairs, among which $N_{+}$ samples own the same emotion label with the input, the precision is defined as:
\begin{equation}
  Precision = \frac{N_{+}}{N}
  \label{eq1}
\end{equation}

We also demonstrate that \sys can generate style embedding closer to ground-truth than the baseline. We calculate the cosine similarities between the generated style embedding and the ground-truth to measure that. 

\input{tables/objective}

As shown in Table \ref{tab:objective}, \sys achieves higher performance than baseline. On the emotional dataset, the precision increases from $0.523$ to $0.776$, while cosine similarity is elevated from $0.749$ to $0.853$. 
On the audiobook dataset, the cosine similarity is elevated from $0.762$ to $0.877$. The objective evaluation metrics indicate that \sys can obtain style embedding more similar to the ground-truth than baseline. 


\begin{figure}[tb]
  \centering
  \includegraphics[width=\linewidth]{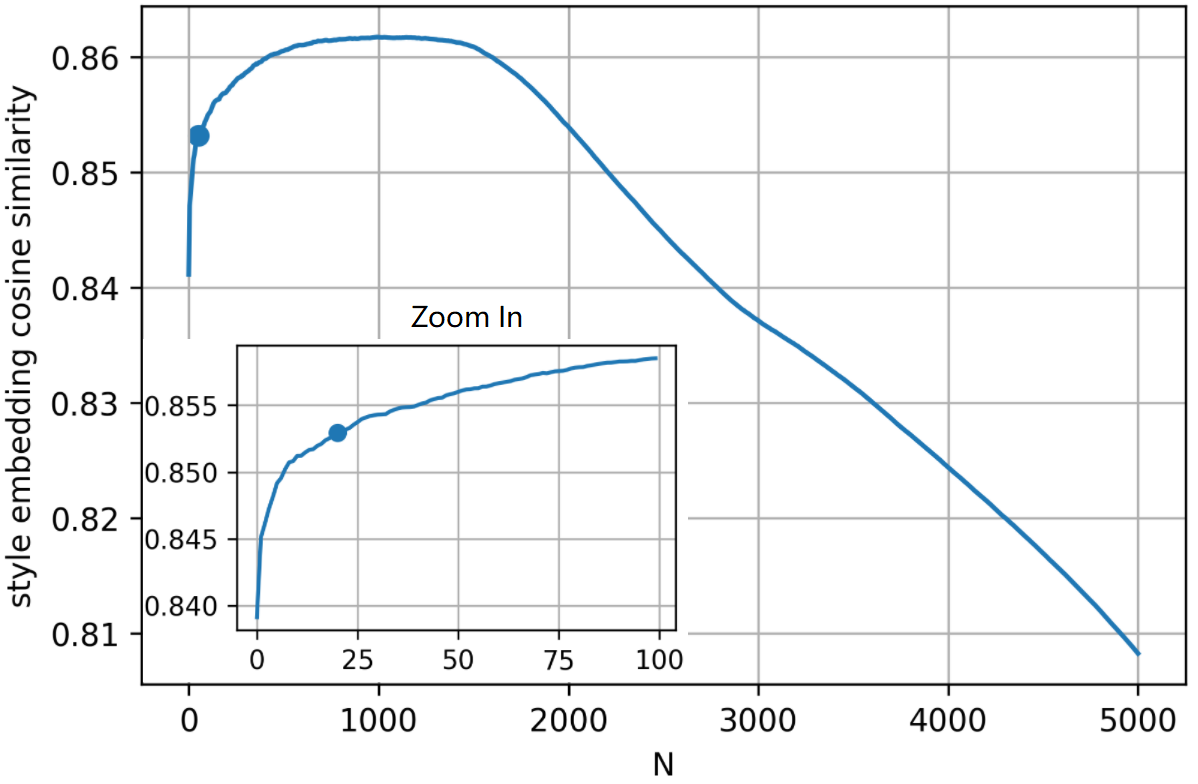}
  \caption{Style similarity with ground-truth style embedding in test set}
  \label{fig:value_N}
\end{figure}

\subsubsection{Subjective evaluation}
We show that \sys also achieves better performance in subjective evaluations.
30 speech files generated by the baseline and proposed approaches are rated by 23 native Mandarin speakers with good English proficiency on a scale from 1 to 5 with 1 point increments according to whether the speaking style is appropriate for the current text, from which a subjective mean opinion score (MOS) is calculated. Meanwhile, the listeners are asked to choose a preferred speech from the speeches synthesized by the baseline and proposed approaches, from which ABX preference rates are calculated.

\input{tables/mos}

As shown in Table \ref{tab:subjective-mos} and \ref{tab:subjective-abx}, the experimental results in both datasets demonstrate the effectiveness of our method. The MOS score increases from $3.160$ to $3.625$ in audiobook dataset and increases from $2.804$ to $3.812$ in emotional dataset. The ABX preference rate exceeds the baseline by $56.7\%$ in emotional dataset and $13.3\%$ in audiobook dataset. 
Both of the subjective evaluations indicate that \sys can generate audios with more appropriate style than baseline.

\subsection{Selection of $N$}
\label{selN}

We investigate the influence of the number of reference audios $N$ in the test set of emotional dataset in Figure~\ref{fig:value_N}.
The result shows that, as $N$ grows, the cosine similarity of the weighted sum style embeddings and the ground-truth style embeddings first increases then decreases. The turning point came around $N $ = 1,500. In the training set, each kind of emotion has 1,450 utterances. 
So we could say the linguistic encoder in CALM extracts text features more related to speaking style.
The similarity increases with $N$, but the computational complexity also increases with $N$. 
Considering the trade-off between performance and complexity, we choose $N$ as $20$, where the improvement of similarity starts to slow down.

%% file: tables/objective.tex
\begin{table}[h]
    \caption{Objective evaluations for different approaches}
    \label{tab:objective}
    \centering
    \begin{tabular}{ccc}
    \toprule
    \textbf{Emotional} & \textbf{Precision} & \textbf{Similarity} \\
    \midrule
    \textbf{Baseline}  & 0.523              & 0.749               \\
    \textbf{Proposed}  & 0.776              & 0.853               \\
    \toprule
    \textbf{Audiobook} & \textbf{Precision} & \textbf{Similarity} \\
    \midrule
    \textbf{Baseline}  & *                  & 0.762               \\
    \textbf{Proposed}  & *                  & 0.877              \\
    \bottomrule
\end{tabular}
\end{table}

%% file: tables/mos.tex

\begin{table}[t]
    \caption{MOS scores with 95\% confidence intervals of the baseline and the proposed
approaches.}
    \label{tab:subjective-mos}
    \centering
    \begin{tabular}{cccc}
        \toprule
        \textbf{MOS}        & \textbf{Baseline}  & \textbf{Proposed} \\
        \midrule
        \textbf{Audiobook}  & $3.160 \pm 0.129$    & $3.625 \pm 0.122$     \\
        \textbf{Emotional}  & $2.804 \pm 0.102$    & $3.812 \pm 0.089$     \\
        \bottomrule
    \end{tabular}
\end{table}

\begin{table}[t]
    \caption{ABX Preference rates between the baseline and the proposed
approaches. NP stands for no preference.}
    \label{tab:subjective-abx}
    \centering
    \begin{tabular}{cccc}
        \toprule
        \textbf{Preference} & \textbf{Baseline} & \textbf{NP} & \textbf{Proposed} \\
        \midrule
        \textbf{Audiobook}  & $30.5\%$            & $25.7\%$     & $43.8\%$            \\
        \textbf{Emotional}  & $16.5\%$            & $10.3\%$      & $73.2\%$           \\
        \bottomrule
    \end{tabular}
\end{table}

  

%% file: sections/conclusion.tex
\section{Conclusion}


We propose a contrastive-learning-based multiple referenced TTS model to synthesize speech with a more appropriate speaking style for the input text.
To demonstrate the effectiveness of our model, we conduct evaluations on two datasets. Our model outperforms the baseline in both objective and subjective evaluation metrics. The experimental results indicate that our model achieves a better style modeling approach and generates more natural audio. 
Finally, we discuss how the number of reference audios $N$ influences the performance and choose it in consideration of both the performance and the computational complexity.

\section{Acknowledgement}
This work is supported by National Natural Science Foundation of China (62076144), National Social Science Foundation of China (13\&ZD189), Shenzhen Key Laboratory of next generation interactive media innovative technology (ZDSYS20210623092001004) and Tsinghua University - Tencent Joint Laboratory.